\documentstyle[twocolumn,prl,aps,psfig]{revtex}
\newcommand{ \be }{\begin{equation}}
\newcommand{ \ee }{\end{equation}}
\newcommand{ \bea }{\begin{eqnarray}}
\newcommand{ \eea }{\end{eqnarray}}
\newcommand{ \la }{\langle}
\newcommand{ \ra }{\rangle}
\newcommand{ \bp }{{\bf p}}

\newcommand{ \br }{{\bf r}}
\newcommand{ \bV }{{\bf V}}
\newcommand{ \bv }{{\bf v}}
\newcommand{ \bk }{{\bf k}}

\begin{document}
\draft
\title{ 
Relative space-time asymmetries in pion and nucleon production
in non-central nucleus-nucleus collisions at high energies}

\author{
  S.~Voloshin$^{1,}$\thanks{On leave from Moscow Engineering Physics
                            Institute, Moscow, 115409, Russia. },
  R.~Lednicky$^2$, S.~Panitkin$^{3,4}$, Nu Xu$^4$
}
\address{
 $^1$ Physikalisches Institut der Universit\"at Heidelberg, Germany
\\ $^2$ Institute of Physics, Prague, Czech Republic
\\ $^3$ Kent State University, USA
\\ $^4$ Nuclear Science Division, LBNL, USA
}

\date{\today}

\maketitle

\begin{abstract}
We propose to use the ratio of the pion-proton correlation functions 
evaluated under different conditions to study 
the relative space-time asymmetries in pion and proton emission
(pion and nucleon source relative shifts) in high energy heavy ion collision.
We address the question of the non-central collisions, where the sources 
can be shifted spatially both in the longitudinal and in the
transverse directions  in the reaction plane.
We use the RQMD event generator to illustrate the effect and the technique.   
\end{abstract}


The importance of the study of the space-time structure of  
particle emission in the heavy ion collisions has been emphasized
repeatedly.
In general the question of the corresponding  measurements
splits into two: the measurements of the extensions (sizes) of the
emission zones of each particular particle species and the measurement of
the relative location of the sources of different particles.
The measurements of  the size of the effective source  could
be addressed in principle by the two (identical) particle
interferometry. At present the particle sources are studied 
intensively by this method.
On the other hand, the question of the relative space and time asymmetry 
in the production of different particles (source shifts) remains 
almost unexplored, 
although this question is very important for an interpretation 
of the experimental data within different models (for example, the
ones which require particle thermalization, chemical equilibration, etc.).

The important first step in the investigation of the source shifts in space 
and time was done in Ref.~\cite{lllen}, where the non-identical particle
correlation functions were proposed as a tool for a study of time delays in 
emission of different particle.
This observation was based on the analytical  calculations~\cite{lll,lll2} 
of the final state interaction contribution to the correlation function.
Recently, more detailed investigation of the possibilities provided by
the correlations of two non-identical particles for the study
of asymmetries in particle production was attempted in~\cite{llpx}.

Here, we use the $\pi^{\pm}p$ correlations to 
study the asymmetries in  the pion and nucleon (proton) production in 
non-central nucleus-nucleus collisions at high energies.
To be specific, we investigate the particle production in 
the rapidity region close
to the projectile rapidity in Au+Au collisions at the AGS energies.
We select the rapidity region where proton directed flow is most
pronounced~\cite{l877flow}, and where it is natural to expect 
that proton and pion sources could be shifted relative to each other
not only in the longitudinal ($z$) direction but also in the flow 
direction (along or opposite to the impact parameter vector,
below, $+x$ or $-x$ directions).

The two-pion interferometry of the non-central collision and its
relation to anisotropic flow was addressed earlier in~\cite{lvc}.
In particular, it was noted that the pion source looks different
from the different directions with respect to the reaction plane angle, 
in part due to the fact that the pion source is
screened by nucleons from the direction of nucleon flow.
It implies that pion source is effectively shifted with respect to 
the proton source. 
This observation was supported~\cite{lvc} by the analysis of the pion 
source  shape in the RQMD generated events.
The recent experimental results on anisotropic flow of pions and 
nucleons~\cite{l877flow} confirm the picture.
The observed flow of protons and pions looks very different, 
and this difference agrees with the assumption of pion screening by nucleons. 
What is even more important that pions of different charge 
at very low transverse momenta flow in the opposite directions.
This fact finds natural explanations in Coulomb interaction with the
protons, assuming that the pion and proton sources are
shifted in space. 
In the current paper we show how such shifts could be detected and 
measured experimentally using pion-proton correlations, determined mainly by
the two-particle final state interaction.


Referring for the quantitative description of the question 
of the final state interaction contribution to the correlation function
to the original papers~\cite{lllen,lll,lll2,llpx}, here we discuss 
it qualitatively.
Such a discussion will help us to explain/justify the techniques
used below.

Any two-particle (both identical and non-identical) correlation
function is mainly sensitive to the distance between 
particles at the moment when the second particle is produced.
Such a separation can be written as 
\be
\tilde{\br}_{12} \approx (\br_1-\br_2) - \bV (t_1-t_2),
\ee
where we introduced the notations $\br$, $t$ for the space and
time position of particle production, and $\bV=(\bp_1+\bp_2)/(E_1+E_2)$
for the pair velocity.
The difference between identical and non-identical particle
correlations (except the quantum statistics effects, which are not
essential for the current study) comes from the fact that the
correlation function in the case of non-identical particles in general
depends on the orientation of the relative velocity 
$\bv \equiv \bp_1/E_1-\bp_2/E_2$ (that is, the correlation function 
is different for the cases of $\bv$ and $-\bv$), 
while for identical particles such a dependence is washed out
due to the symmetry of the system.

The dependence of the non-identical two-particle correlation function
on the relative velocity becomes clear if one recollects
that the non-identical two-particle correlations are mostly due to
the two body interaction in the final state. 
The impact of the two body final state interaction obviously is different 
for particles (at the instant of both of them are created) 
moving, in the particle pair rest frame, 
toward each other or moving in the opposite directions.
If the centers of the effective sources of two particles are shifted,
then the dependence of the correlation function on the orientation of
the relative velocity with respect to the shift becomes obvious. 
Suppose that the particles of type ``2'' are produced on average at 
$z_2 >z_1$ (for simplicity assume also that  $t_2 \approx t_1$). 
Then, selecting the pairs with $v_{1,z} > v_{2,z}$ one would select 
the pairs which move on average toward each 
other (in the  pair rest frame) and the corresponding
final state interaction would be stronger.
(In this simplified consideration we also neglect the possible differences
between particle velocities at the moment of the production and the measured 
ones). 
Depending on the type of final state interaction 
the correlation function would decrease or increase.
Studying the orientation of the relative velocity to which the
correlation function is the most sensitive one could find the orientation
of the shift in space. 

It is natural to study the two non-identical particle ($\pi^{\pm}p$)
correlations in the pair rest frame and as a function 
of $\bk^*$, the (half of) particle relative momentum 
in this system ($\bk^*=\bp_{\pi}=-\bp_{p}$).
In the region of small values of $k^*$ (much less than the inverse 
size of the particle source) the final state interaction 
of two charged particle is dominated by Coulomb interaction. 
In this case the correlation function can be written in the analytical 
form~\cite{lll,lll2,llpx},
which makes it possible to estimate the ratio 
of the correlation functions evaluated under different conditions.
If we denote the correlation functions for 
two different cases, for example, $k^*_i>0$ 
and  $k^*_i<0$,
as $R^{(+)}_i$ and $R^{(-)}_i$,
then in the limit of small $k^*$, 
\bea
\frac{R^{(+)}_i}{ R^{(-)}_i }
&\approx&
\frac{a + 2\la r^* \ra +2 \la \br^*  \bk^*/k^*\ra^{(+)}}
{a + 2\la r^* \ra +2 \la \br^* \bk^*/k^* \ra^{(-)}}   
\nonumber          \\
&\approx&
\frac{1 + 2 \la \br^*\ra \la \bk^*/k^*\ra^{(+)}/a}
{1 + 2 \la \br^* \ra  \la \bk^*/k^* \ra^{(-)}/a}
\approx 1+2 \la \br^* \ra_i /a,
\label{erprm}
\eea
where $a$ is the Bohr radius (for the $\pi^\pm p$ system 
$a \approx\pm 222$~fm), $\br^*$ is the relative separation in particle 
production points.
Then,  $\la \br^* \ra$ is the shift between sources in the pair rest frame,
the quantity of interest.
It should be mentioned, that in the derivation of Eq.~(\ref{erprm})
it is assumed that $\la r^* \ra \ll |a|$, 
and that $|{\rm Re} f| \ll |a|$, where $f$ is 
the strong s-wave pion-proton scattering amplitude.
The average in formula (\ref{erprm}) is taken over all pairs from
the corresponding subsample. 
It was also used that in the limit of $k^* \ll \la p_t\ra$,
$ \la \br^* \ra \la \bk^*/k^*\ra^{(\pm)} = \pm \la \br^* \ra_i/2$ 
for the above defined cuts on $k^*_i$.


For the current study we use the RQMD v1.08~\cite{lrqmd} 
event generator to simulate Au+Au collision at 11.4~GeV/nucleon.
We select {\em particles} in the rapidity region $2.8<y_{lab}<3.2$ 
and within relatively narrow sector in the azimuthal space,
always such that $p_x>0$ and $|p_y/p_x| <0.5$.
We create two event subsamples in accordance to 
the orientation of the reaction plane, ``$\Psi_r=0$''
(in this case nucleons flow in the positive $x$ direction) and
``$\Psi_r=\pi$'' subsamples.
The relative shifts between pion and proton sources for both cases
are presented in Table~1 as calculated both in the center of mass
system of colliding nuclei and in the particle pair rest frame.
In the RQMD generated events we observe strong momentum-position 
correlations in particle production (especially for pions); 
the values presented in the table
correspond to the average of the production points over the particles 
which contribute to the pairs with $k^* < 15$~MeV.
This restriction limits significantly the effective transverse
momentum range of pions, 
decreasing its average value in the sample. 
For both subsamples the average components of the pair velocity 
in the center of mass system of colliding nuclei are
$V_z \approx 0.89$ and $V_x \approx 0.17$.

We analyze approximately 200K pairs in each of the subsamples, 
the typical statistics for modern experiments.
Our goal is to show the sensitivity of the $\pi p$ correlation
function to the shifts presented in Table~1


Below we consider only $\pi^+ p$ correlation function (which {\em
decreases} with the strength of the interaction),
the results for $\pi^- p$ case would be very similar taking into
account the change in sign of Coulomb interaction, which
dominates at low values of particle relative velocity.
The correlation functions are calculated in accordance to 
Lednicky-Lyuboshitz~\cite{lll} formulae taking into account both
Coulomb as well as the strong interaction in the final state. 
In Fig.~1 we show the correlation functions calculated for the ``$\Psi_r=0$''
event subsample  for two sets of cuts $k^*_x>0$ ($k^*_x<0$)
and  $k^*_z>0$ ($k^*_z<0$). 
Note that these cuts approximately correspond to the cuts in 
the laboratory system
$v_{\pi,x} > v_{p,x}$ ($v_{\pi,x}<v_{p,x}$) and $v_{\pi,z}>v_{p,z}$ 
($v_{\pi,z}<v_{p,z}$),
respectively. The correspondence would be strict if one selects the coordinate
system with one of the axes parallel to the pair velocity.
The ratios of the correlation functions are also shown on the same figure.
One can see that the correlations are stronger in the cases of
$k^*_z>0$ and $k^*_x>0$, which clearly indicates that the proton
source is shifted relative to the pion source 
to positive $z^*$ and positive $x^*$ values. 
The magnitude of the difference in the cases of cuts on $k^*_z$ and
$k^*_x$ shows that the shift in $z$ direction is larger than the one
in $x$ direction.

In Fig.~2 we show the analogous plots for the ``$\Psi_r=\pi$'' subsample.
Note the difference in Fig.~1 an Fig.~2 concerning the cut on $k^*_x$.
It reflects the fact that the relative shift between pion and proton
sources has changed the sign (and its value), and the proton source 
is now to the ``left'' relative to the pion source.
The changes in the correlation function (``$k^*_x$'' cuts)
although small in magnitude are significant statistically, 
what is seen in Fig.~3, where we show
the correlation functions  at a different scale.


Qualitatively the results of the $\pi^+ p$ correlation analysis show
unambiguously the correct relative location of pion and proton sources.
Quantitatively the agreement with the approximate
formula~(\ref{erprm}) is also very good. 
In accordance to (\ref{erprm}), each one fermi of the shift leads
to the change of approximately 0.9\% in the ratio of the corresponding
correlation functions at small values of $k^*$.
If one compares the numbers from the Table I with the values
of the correlation function ratios extrapolated to $k^*=0$ from Figs.~1--3, 
a good quantitative agreement will be found.
From the Table I one would expect the following values for the ratios 
of the correlation functions at small values of $k^*$:
in the case of ``$\Psi_r=0$'' the ratio 
$R^{(-)}_z/R^{(+)}_z \approx 1.11$ and $R^{(-)}_x/R^{(+)}_x \approx 1.05$;
in the case of ``$\Psi_r=\pi$'',
$R^{(-)}_z/R^{(+)}_z \approx 1.09$ and $R^{(-)}_x/R^{(+)}_x \approx 0.99$.
The correlation function ratios 
shown in Fig.~1--3 agree with these numbers quite well.
 
In the current paper we have analyzed the correlation function in the
particle pair rest frame. 
The source shifts which were extracted are the ones in this frame. 
Although, in principle, the correlation function is sensitive 
to the time shift between the sources~\cite{lll,lll2} in this system, 
the sensitivity is  very weak and can be neglected under the condition of 
$|t^*| \ll \mu {r^*}^2$ ($\mu$ is the reduced mass). 
Physically, one of the reasons for this is the very small particle relative
velocity in this frame $\bv^* \approx \bk^*/\mu$; 
then the relative space separation between particles changes on average very
little  during the time $\Delta t=t^*$ provided that 
$k^*/\mu |\Delta t| \ll | \la \br^* \ra |$.

If one considers the space and time shifts in the laboratory system 
(or in the center of mass of colliding nuclei system) then the correlation
function depends on all four shifts, but all four of them cannot be 
extracted from 3 possibly measured ratios of the correlation functions.  
It is exactly the same problem as observed in the two-particle interferometry. 
To extract all shifts in this case  one needs to make additional assumptions, 
for example, that the shifts do not depend on the velocity of the pair.
Such questions are beyond the scope of the current publications.


We have shown that the important information on the relative shifts 
between pion and proton sources can be obtained by comparing the $\pi p$
correlation functions evaluated at different conditions.
The good {\em quantitative} results achieved 
in the application of the method to the RQMD generated
events give a hope that the same technique can be successfully 
applied to the data.


Two of the authors (S.V. and R.L.) wish to acknowledge the support 
and hospitality of the Nuclear Theory and the Relativistic Nuclear Collisions
groups of the Nuclear Science Division of the Lawrence Berkeley National
Laboratory during their visit to LBNL. 
This work was supported in part by the US DoE
Contract No.DE-AC03-76SF00098 
and by GA AV Czech Republic, Grant No. 1010601.


\begin{table}[htbp]
\caption[]{\footnotesize
The mean values of spatial and temporal shifts (in~fm) 
of pion and proton sources for two different orientations of the 
reaction plane in the system of the center of mass of colliding nuclei 
(upper half) and in the pair rest frame (lower half).
}
\label{trq}

\medskip
\begin{center}
\begin{tabular}{|c|c|c|c|c|} 
\hline
     &$\la x_{\pi}-x_{p} \ra$
     &$\la y_{\pi}-y_{p} \ra$
     &$\la z_{\pi}-z_{p} \ra$
     &$\la t_{\pi}-t_{p} \ra$ \\
\hline
$\Psi_r=0  $ &   -4.7   &  0.1     &  -8.3    &  -3.7     \\
$\Psi_r=\pi$ &    1.5   &  0.1     &  -7.1    &  -2.8     \\
\hline
\hline
     &$\la x_{\pi}^*-x_{p}^* \ra$
     &$\la y_{\pi}^*-y_{p}^* \ra$
     &$\la z_{\pi}^*-z_{p}^* \ra$
     &$\la t_{\pi}^*-t_{p}^* \ra$ \\
\hline
$\Psi_r=0  $ &   -5.8   &  0.1     &  -12.3   &  10.3    \\
$\Psi_r=\pi$ &    0.9   &  0.2     &  -10.5   &   8.3    \\
\hline
\end{tabular}
\end{center}
\end{table}

\begin{figure}
\centerline{\psfig{figure=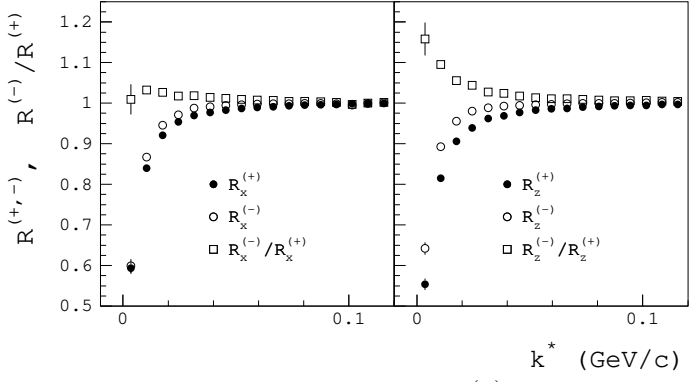,height=5.0cm}}
  \caption[]{
The correlation functions $R^{(+)}_{x,z}$ ($k^*_{x,z} >0  $) 
and $R^{(-)}_{x,z}$ ($k^*_{x,z} <0 $) for the event 
subsample ``$\Psi_r=0$''.
    }
\label{fig1}
\end{figure}

\begin{figure}
\centerline{\psfig{figure=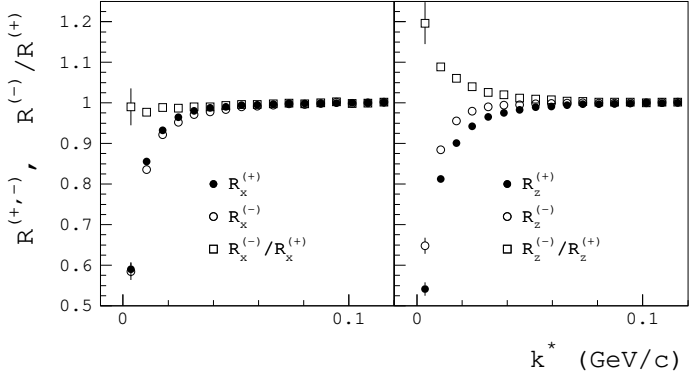,height=5.0cm}}
  \caption[]{
The same as Fig.~1 for the event subsample ``$\Psi_r=\pi$''.
    }
\label{fig2}
\end{figure}

\begin{figure}
\centerline{\psfig{figure=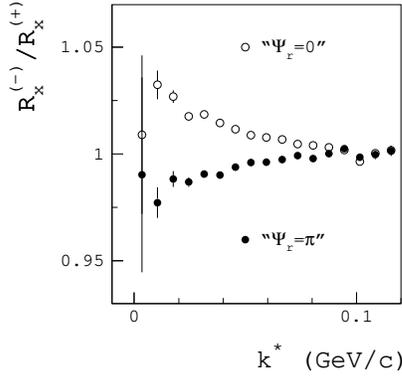,height=5.0cm}}
  \caption[]{
The ratios of the correlation function $R^{(-)}_x$ 
and $R^{(+)}_x$  for two event subsamples ``$\Psi_r=0$''
and ``$\Psi_r=\pi$''.
    }
\label{fig3}
\end{figure}

\end{document}